\documentclass[aps,prd,showpacs,groupedaddress,floatfix,nofootinbib]{revtex4}
\usepackage{amsmath}
\usepackage{graphicx}
\usepackage{color}

\newcommand{\beq}{\begin{eqnarray}}
\newcommand{\eeq}{\end{eqnarray}}

\begin{document}

\title{{Comment on ``Quantization of FRW spacetimes \\ in the presence of a cosmological constant and radiation''}}
\author{Paolo Amore}
\email{paolo@ucol.mx}
\affiliation{Facultad de Ciencias, Universidad de Colima,\\
Bernal D\'{i}az del Castillo 340, Colima, Colima, M\'exico.}
\author{Alfredo Aranda}
\email{fefo@ucol.mx}
\affiliation{Facultad de Ciencias, Universidad de Colima,\\
Bernal D\'{i}az del Castillo 340, Colima, Colima, M\'exico.}
\author{Mayra Cervantes}
\email{mayradcv@ucol.mx}
\affiliation{Facultad de Ciencias, Universidad de Colima,\\
Bernal D\'{i}az del Castillo 340, Colima, Colima, M\'exico.}
\author{J. L. D\'iaz-Cruz}
\email{jldiaz@fcfm.buap.mx}
\affiliation{Facultad de Ciencias, Universidad de Colima,\\
Bernal D\'{i}az del Castillo 340, Colima, Colima, M\'exico}
\affiliation{Facultad de Ciencias F\'{i}sico-Matem\'aticas, BUAP \\
Apdo. Postal 1364, C.P.72000 Puebla, Pue, M\'exico}
\author{Francisco M. Fern\'andez}\email{fernande@quimica.unlp.edu.ar}
\affiliation{INIFTA (Conicet,UNLP), Divisi\'on Qu\'imica Te\'orica, Diag. 113 y 64 S/N, \\
Sucursal 4, Casilla de Correo 16, 1900 La Plata, Argentina}

\begin{abstract}
The quantization of the Friedmann-Robertson-Walker
spacetime in the presence of a negative cosmological constant was used in a recent paper
to conclude that there are solutions that avoid singularities (big bang--big crunch) at 
the quantum level. We show that a proper study of their model does not indicate the it 
prevents the occurrence of singularities at the quantum level, in fact the quantum probability 
of such event is larger than the classical one. 
Our numerical simulations based on the powerful variational sinc collocation method (VSCM) also show that the
precision of the results of that paper is much lower than the $20$ significant digits reported by the authors.
\end{abstract}


\maketitle

We review the results presented in a recent paper on Quantum Cosmology by
Monerat et al~\cite{MCOFL06},
concerning the quantization of the Friedmann-Robertson-Walker spacetime. The quantum dynamics of this model
is described by the Wheeler-De Witt equation, which is obtained by applying the Dirac formalism
to the quantization of constrained systems. Such equation reads
\beq
\left(\frac{1}{12} \frac{\partial^2}{\partial a^2}  - 3 k a^2 + \Lambda a^4 \right) \Psi(a,\tau) =
- i \frac{\partial}{\partial \tau} \Psi(a,\tau)  \ ,
\label{WdW}
\eeq
and it is formally equivalent to the Schr\"odinger equation for a quartic anharmonic
oscillator with a potential given by~\cite{MCOFL06}
\beq
V_e(a) = 36 k a^2 - 12 \Lambda a^4 \ .
\label{eq:Ve}
\eeq
Here $a$ is the scale factor in the FRW metric, $\Lambda$ is the (negative) cosmological constant
and $k$ denotes the constant curvature of the spatial sections.

The corresponding time--independent Schr\"odinger equation then reads
\beq
\left( - \frac{\partial^2}{\partial a^2}  + V_e(a) \right) \eta(a) = 12 E \eta(a) ,
\eeq
in the notation of Monerat et al~\cite{MCOFL06}. In what follows we write $\varepsilon \equiv 12 \ E$.

Our critique of the arguments of Monerat et al~\cite{MCOFL06} is based on both technical
and conceptual grounds. The former stems from the numerical results obtained in their paper which are below the
claimed accuracy. The latter concerns the correct interpretation of the model
which does not support the conclusion that singularities are avoided at a
quantum level~\cite{MCOFL06}.

Regarding the technical part, Monerat et al~\cite{MCOFL06} obtained their results by
means of a technique developed by Chhajlany and Malnev~\cite{CM90,CM91}. The application of this
method requires the introduction of an extra term in the Schr\"odinger equation, in the form of
a sextic potential with a coefficient obtained from a cubic algebraic equation.
Due to this fact the energies and wave functions calculated by Monerat et al are only
approximations with roughly $1\%$ error. However, the results
reported in Table I of  Ref.~\cite{MCOFL06}, which are obtained using the modified
potentials given in their equations (26), (30) and (32) are claimed to have $20$ digits
accuracy.
Long time ago Fern\'andez~\cite{FMF91} noticed that the method of
Chhajlany et al~\cite{CM90,CM91} converges rather slowly
and it therefore requires a much larger number of terms to reach that precision.

In order to verify the accuracy  of the results we have used the powerful
``Variational Sinc Collocation Method'' (VSCM)~\cite{Amore06a,Amore06b}, which
allows one to obtain very precise numerical results,
both for the energies and wavefunctions, and has proved to yield errors
that decay exponentially with the number of elements (sinc functions) used.

In what follows we briefly sketch the VSCM. The sinc functions are given by
\beq
S_k(h,x) &\equiv& \frac{\sin\left(\pi (x-k h)/h\right)}{\pi (x-k h)/h}  \ ,
\eeq
where $k = 0, \pm 1,\dots$ is an integer and $h$ is the spacing between two
contiguous sinc functions.
The sinc functions are orthogonal:
\beq
\int_{-\infty}^{+\infty} S_k(h,x) \ S_l(h,x) \ dx = h \ \delta_{kl}  .
\eeq

Using a collocation scheme, one can express the matrix element of the Hamiltonian operator in the
sinc ``basis'' as
\beq
H_{kl} \approx h \left[- c_{kl}^{(2)} + \delta_{kl} \ V(k h)  \right] \ ,
\eeq
where
\beq
c_{lk}^{(2)} &\equiv& \left\{ \begin{array}{c}
- \frac{\pi^2}{3 h^2} \ \ if \ \ k=l \\
- \frac{2}{h^2} \frac{(-1)^{k-l}}{(k-l)^2} \ \ if \ \ k \neq l
\end{array}
\right.
\eeq
are the coefficients obtained from the discretization of the second derivative.
Diagonalization of the $N \times N$ matrix yields a set of $N$ eigenvalues (energies)
 and eigenvectors (wavefunctions). Notice however, that the diagonalization of $H$
 requires the specification of the otherwise arbitrary grid spacing $h$.
 For a given number $N$ of sinc functions there exists
an optimal value of $h$ which provides the smallest errors.
As shown by Amore and collaborators~\cite{Amore06a,Amore06b}
such optimal value of $h$ can be found by application of the Principle of Minimal
Sensitivity (PMS)~\cite{Ste81}
to the trace of the $N \times N$ hamiltonian matrix. Notice that once the hamiltonian
matrix has been
diagonalized one can also solve the time--dependent Schr\"odinger equation,
in terms of the stationary states previously calculated.

Our table \ref{tab1} shows the odd energies for the
potentials given in equations (26), (30) and (32) of  Ref.~\cite{MCOFL06} using the VSCM.
The results in this table are correct up to the $20^{th}$ digit, and show that the results 
reported by Monerat et al in their Table I \cite{MCOFL06} are considerably less precise 
than they appear to suggest.
Just to mention one example,
their energy $E_{16}$ corresponding to $k=1$
has just $5$ correct digits. Although this lack of precision does not affect the
authors' analysis~\cite{MCOFL06},
which is in any case bounded by the more severe $1\%$ error due to the use of an
effective potential rather than the
exact one, we believe that it is important to remark this point.

\begin{table}
\caption{\label{tab1} Lowest energy levels for the effective potentials with $k=0,\pm 1$. }
\begin{ruledtabular}
\begin{tabular}{c|ccc}
Level & $k = 1$ & $k = 0$  & $k = -1$ \\
\hline
 $E_1$ &  $1.5103016760578707333$   & $0.3404027974289474964$ & $-10.553138375809812184$\\
 $E_2$ &  $3.5509871014722926520$   & $1.0496018816142096406$ & $-8.8509608822646650503$ \\
 $E_3$ &  $5.6230931685079239340$   & $1.9248782225982734499$ & $-7.2247835560429854525$\\
 $E_4$ &  $7.7256531719676264015$   & $2.9232145294809850623$ & $-5.6802949920840011967$\\
 $E_5$ &  $9.8577817763092210058$   & $4.0227383298745516995$ & $-4.2251989178977204001$\\
 $E_6$ &  $12.018664366203979043$   & $5.2097960735587976352$ & $-2.8707515711793779600$\\
 $E_7$ &  $14.207548236413247839$   & $6.4748734068112171253$ & $-1.6348253113363418720$\\
 $E_8$ &  $16.423735233636153025$   & $7.8108774875618647285$ & $-0.5364307199618285749$\\
 $E_9$ &  $18.666575557925837760$   & $9.2122719234208330574$ & $0.481759254997070592451$\\
$E_{10}$& $20.935462499941518896$   & $10.674588220211030332$ & $1.583190805382159479497$\\
$E_{11}$& $23.229827940734217332$   & $12.194126732921328651$ & $2.833031712002511140698$ \\
$E_{12}$& $25.549138478317595566$   & $13.767762283584094708$ & $4.213983702088123106309$\\
$E_{13}$& $27.892892073596735435$   & $15.392811775837602761$ & $5.707834132050646278314$\\
$E_{14}$& $30.260615129836301427$   & $17.066940597587263816$ & $7.302868225546844668469$ \\
$E_{15}$& $32.651859936513780000$   & $18.788094386112582846$ & $8.990905434198164180404$ \\
$E_{16}$& $35.066202421382313215$   & $20.554447991460572182$ & $10.76576537713456435413$\\
$E_{17}$& $37.503240164769059826$   & $22.364366463785281446$ & $12.62252647249543208197$\\
$E_{18}$& $39.962590638221848703$   & $24.216374668801875820$ & $14.55712323817216360861$\\
$E_{19}$& $42.443889636078437368$   & $26.109133235245860851$ & $16.56610649829916492072$\\
$E_{20}$& $44.946789873733933173$   & $28.041419241199542663$ & $18.64649041909474222481$\\
$E_{21}$& $47.470959730597782399$   & $30.012110508653406479$ & $20.79564954975016308111$\\
$E_{22}$& $50.016082119171135490$   & $32.020172687758339925$ & $23.01124646297271199397$\\
$E_{23}$& $52.581853464498535137$   & $34.064648527576543759$ & $25.29117906504417774373$\\
\end{tabular}
\end{ruledtabular}
\bigskip\bigskip
\end{table}

\begin{table}
\caption{\label{tab2} Lowest energy levels for the exact potential with $k=0,\pm 1$. }
\begin{ruledtabular}
\begin{tabular}{c|ccc}
Level & $k = 1$ & $k = 0$  & $k = -1$ \\
\hline
 $E_1$ &     $1.5102625384705021518$  &  $0.3364795921008755186$  &   $-21.79569603947057798056$ \\
 $E_2$ &     $3.5506472908620463050$  &  $1.0311990503013958552$   &  $-20.39848968823294656013$\\
 $E_3$ &     $5.6218937064348010503$  &  $1.8807615814216099958$  &   $-19.01885125967730517826$\\
 $E_4$ &     $7.7227778139934265020$  &  $2.8424874930254863943$  &   $-17.65745105165687339198$\\
 $E_5$ &     $9.8521942203839334149$  &  $3.8947462112216761884$   &  $-16.31503013488920386315$\\
 $E_6$ &     $12.009138568080285499$  &  $5.0240915609641800702$  &   $-14.99241327017434870429$\\
 $E_7$ &     $14.192693387434484198$  &  $6.2211924298081547999$   &  $-13.69052532294706014390$\\
 $E_8$ &     $16.402016547726087957$  &  $7.4791208557098708821$  &   $-12.41041248037188872561$\\
 $E_9$ &     $18.636331727173164959$ &   $8.7924907241645713076$   &  $-11.15327022078174141020$\\
 $E_{10}$ &  $20.894920472196314023$  &  $10.156971843165374589$   &  $-9.920481041305256697868$\\
 $E_{11}$ &  $23.177115522582541829$  &  $11.568992820956286787$   &  $-8.713666752230586250096$\\
 $E_{12}$ &  $25.482295155850129982$  &  $13.025548149708503746$ &    $-7.534763378250629251892$\\
 $E_{13}$ &  $27.809878360230719403$  &  $14.524066967082505523$   &  $-6.386132854421764354442$\\
 $E_{14}$ &  $30.159320687349375640$ &   $16.062320376957969417$   &  $-5.270738293115111860300$\\
 $E_{15}$ &  $32.530110667020528246$  &  $17.638353963148189178$  &   $-4.192437789021558602284$\\
 $E_{16}$ &  $34.921766690427654626$  &  $19.250437372785090866$  &   $-3.156519423136907030485$\\
 $E_{17}$ &  $37.333834286304756785$  &  $20.897025823964631886$  &   $-2.170719224552462670164$\\
 $E_{18}$ &  $39.765883728998739663$  &  $22.576730162849598220$  &   $-1.246288323062400948910$\\
 $E_{19}$ &  $42.217507928478425449$  &  $24.288293189747778350$  &   $-0.389996330189812857692$\\
 $E_{20}$ &  $44.688320561206965683$  &  $26.030570672865707451$  &   $0.433719867112653240009$\\
 $E_{21}$ &  $47.177954407853964266$  &  $27.802515928224750562$  &   $1.300741393435948191383$\\
 $E_{22}$ &  $49.686059869496678687$  &  $29.603167154324565317$   &  $2.243855389930394850590$\\
 $E_{23}$ &  $52.212303638550624930$  &  $31.431636924008847119$  &   $3.256001768433707607172$\\
\end{tabular}
\end{ruledtabular}
\bigskip\bigskip
\end{table}

We wish to stress that the numerical method that we have used is completely general and can be
applied to arbitrary potentials; for this reason we have decided to perform the numerical simulation
and time evolution of the model with the exact potential, rather than following the strategy of
 Monerat et al~\cite{MCOFL06}, which was dictated by their choice of numerical method~\cite{CM90,CM91}.
Table \ref{tab2} shows the eigenvalues corresponding to the exact potential Eq.~(\ref{eq:Ve}).
Particularly in the case $k=-1$, corresponding to a double well we have verified that the results
are quite different from the ones obtained with the effective potential (compare the fourth
columns of tables \ref{tab1} and \ref{tab2}).

By means of the VSCM we have obtained accurate energies $E_n$ and
wavefunctions $\phi_n(a)$ up to $n=23$.
Using these states we can  build a wave packet
\beq
\Theta(a,\tau) = \sum_n B_n \phi_n(a) \ e^{-i \varepsilon_n \tau}
\label{wp}
\eeq
where the coefficients $B_n$ are taken to be equal to unity as in Ref.~\cite{MCOFL06} and the energies 
$\varepsilon_n = 12 E_n$. The expected value of the scale factor is 
\beq
\langle a \rangle = \frac{\int_0^\infty  a \ |\Theta(a,\tau)|^2 da}{\int_0^\infty |\Theta(a,\tau)|^2 da} =
\frac{\sum_{nn'} B_n^* B_{n'} \kappa_{nn'} e^{ i (\varepsilon_n-\varepsilon_{n'})\tau}  }{\sum_n |B_n|^2 } ,
\eeq
where $\kappa_{nn'} \equiv \int_0^\infty a \ \phi_n^*(a) \phi_{n'}(a) da$.
This expectation value oscillates in time because of the time--dependent terms
$ e^{ i (\varepsilon_n-\varepsilon_{n'})t}$. Likewise the energy of the wave packet is
${\cal E}_{wp} = \sum_n |B_n|^2 \varepsilon_n /\sum_n |B_n|^2$. 

\begin{figure}
\begin{center}
\includegraphics[width=15cm]{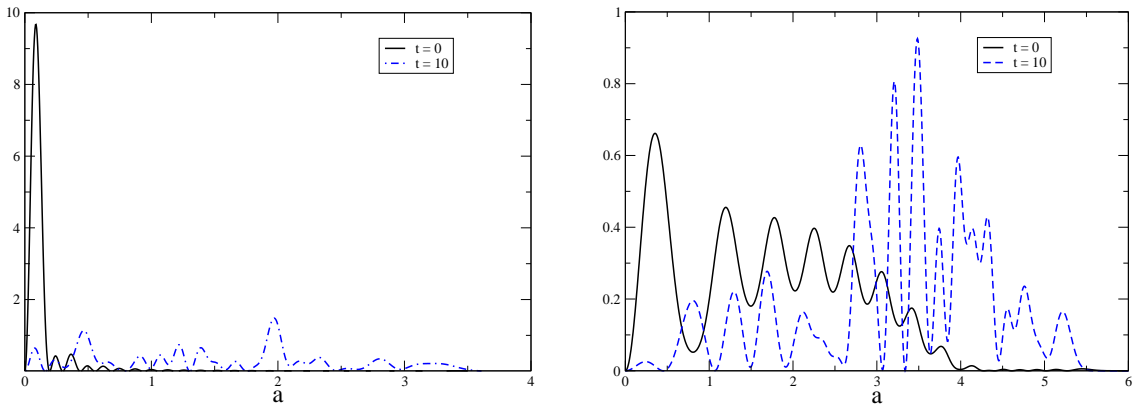}
\bigskip
\caption{Wave packets for $\Lambda = -0.1$ and $k=1$ (left plot) and $k=-1$ (right plot) at two different times. (color online)}
\bigskip
\label{FIG2}
\end{center}
\end{figure}

We notice that the wave packet of eq.~(\ref{wp}) is not yet fully specified, since the
eigenfunctions $\phi_n(a)$ are known up to a phase factor which can be chosen arbitrarily. 
By looking at the coefficients in Table II of \cite{MCOFL06} and of Tables II-VII of 
\cite{MCOFL06arxiv} we see that these wave functions have been chosen by Monerat and 
collaborators to be positive in a small neighborhood of $a=0^+$. The wave packet obtained with this 
prescription for $k=1$ corresponds at $t=0$ to a state sharply localized around $a=0.1$, which quickly spreads at
later times according to the Heisenberg principle. The reader can look at the left plot in Fig.~\ref{FIG2}, where the
wave packet at $t=0$ and $t=10$ is plotted . The wave packet for  $k=-1$ is plotted in the right plot of Fig.~\ref{FIG2}.

\begin{figure}
\begin{center}
\includegraphics[width=15cm]{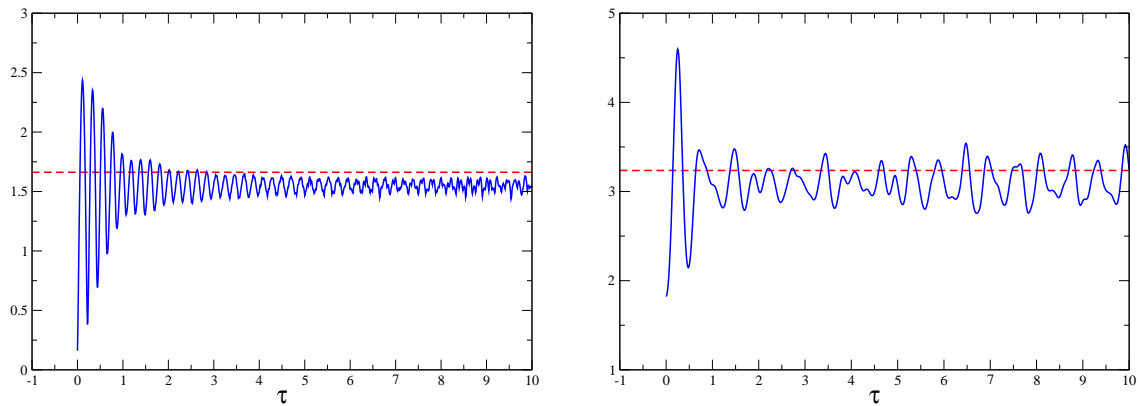}
\bigskip
\caption{Behavior of the expectation value $\langle a\rangle$ for $\Lambda = -0.1$ and 
$k=1$ (left plot) and $k=-1$ (right plot). (color online)}
\bigskip
\label{FIG1}
\end{center}
\end{figure}

Fig.1 of \cite{MCOFL06} displays the expectation value of the scale factor for $\Lambda=-0.1$ and $k=1$; in this plot
it is difficult to appreciate the localization of the wave packet, since the curve is compressed by the large time 
scale considered. In fig.\ref{FIG1} of the present paper we show the behavior  of the expectation value 
$\langle a\rangle$ for $\Lambda = -0.1$ and $k=\pm 1$ using the same choice of phases of Monerat et al. 
Our result for $k=1$ is similar to the one observed by Monerat et al.: the expectation value of the scale factor starts at $a\approx 0.1$ at
$t=0$ and then strongly oscillates at later times, in accordance with the Heisenberg principle; for $k = -1$, however, our curve is 
different from the corresponding curve of Fig.3 of  \cite{MCOFL06}, and $\langle a \rangle$ oscillates around 
a larger value. The reason for this discrepancy is the poor accuracy of the results of Monerat et al. especially for the 
case $k=-1$: the fact that $\langle a \rangle$ oscillates around $a \approx 3$ in our figure reflects both the incorrectness of the
eigenfunctions calculated in \cite{MCOFL06} and the presence of a minimum of the potential at $a=\sqrt{\frac{3 k}{2\Lambda}} = \sqrt{15}$ 
(being the energy of the wave packet negative).

The dashed horizontal lines in our figure correspond to the classical expectation value  $\langle a\rangle_{cl}$
for a particle moving in the same potential and with the same energy of the wave packet.
The classical probability of finding a given value for the scale factor is given by
\beq
P_{cl}(a) = \frac{N_{cl}}{\sqrt{V(a_+)-V(a)}}
\eeq
where $N_{cl} = 1/\int_0^{a_+} \frac{da}{\sqrt{V(a_+)-V(a)}}$ 
and $a_+$ is the right inversion point (remember that in our problem there is an infinite
wall at the origin) given by ${\cal E}_{wp} = V(a_+)$.

Based on the behavior of $\langle a\rangle$ in their figures 1-3 the authors of \cite{MCOFL06} 
state that they ``may be confident about all values assumed by $\langle a \rangle$, in particular, 
that it does not vanish'', thus concluding that this proves the absence of a big-crunch in the 
model at the quantum level or  using their own words ``since the expectation values of the scale 
factors never vanish, we have an initial indication that these models may not have singularities 
at the quantum level''.

In our opinion these observations are not correct: in the first place, one concludes that 
$\langle a \rangle \neq 0$ {\sl without} any calculation because it is a consequence of the 
presence of an infinite wall located at $a=0$ and {\sl is therefore completely independent 
of the form of the potential}.
In the second place, and more importantly, the condition  $\langle a \rangle \neq 0$
does not imply the absence of a big crunch and actually {\sl holds even at the classical level}. 
As a matter of fact, simple mathematical considerations, based on the form of $P_{cl}(a)$, 
are sufficient to conclude that $\langle a \rangle_{cl} >0$~\footnote{The quantum and 
classical expectation values are obtained as integrals of a positive definite function 
and therefore it can {\sl never} vanish.}. 
As we have said the horizontal curves in Fig. \ref{FIG1} correspond to the classical expectation 
values which in both cases fall in the region of  oscillation of the quantum expectation value.
For this reason we believe that hardly any conclusion can be drawn from the study of $\langle a \rangle$
performed in Fig.1-3 of \cite{MCOFL06}.

We also stress that, although the expectation value of the scale factor can never vanish, 
one can build wave packets which at a given time (say $t=0$) are localized around an 
arbitrarily small values, $a_0$. 
Such wave packets are obtained using $B_n = \phi_n^*(a_0)$ for the first $N$ states of the potential 
using the completeness of the basis $\left\{ \phi_n(a)\right\}$:
\beq
\delta(a-a_0) \approx \sum_{n=0}^N \phi^*_n(a_0) \phi_n(a) \ .
\eeq

\begin{figure}
\begin{center}
\includegraphics[width=15cm]{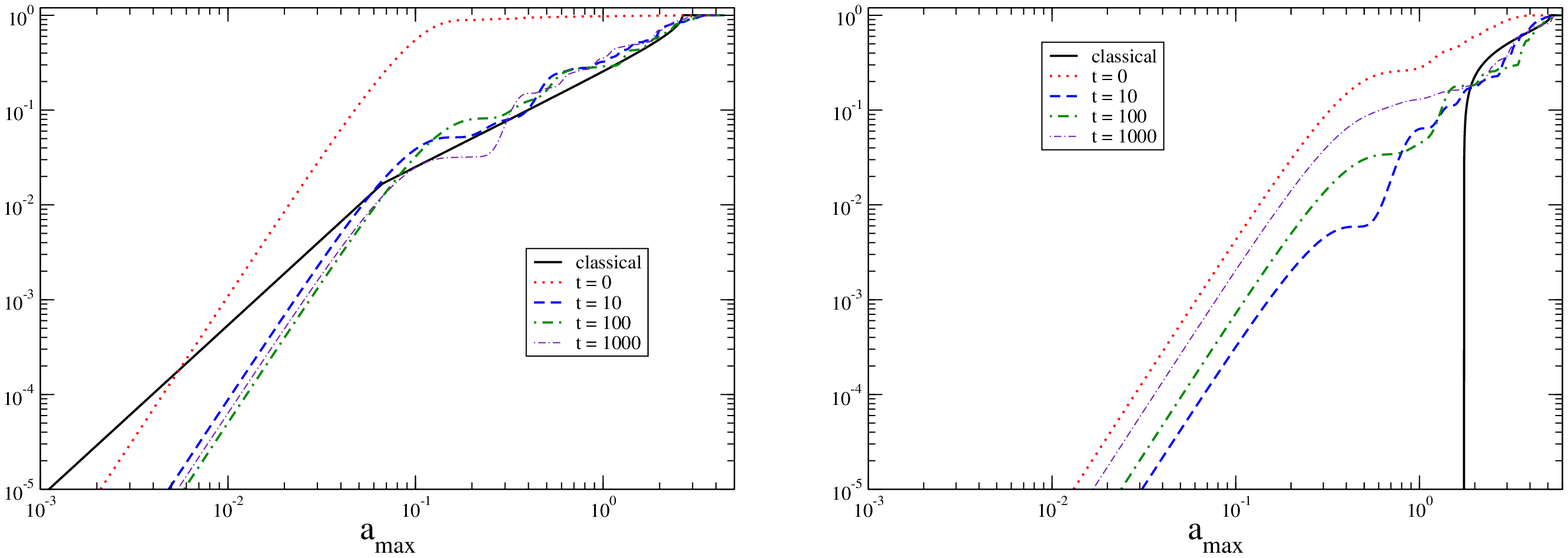}
\bigskip
\caption{Probability of finding a scale factor $\leq a$  for $\Lambda = -0.1$ and $k=1$ (left plot) and 
$k=-1$ (right plot). (color online)}
\bigskip
\label{FIG3}
\end{center}
\end{figure}

On the other hand, a more meaningful comparison between the classical and quantum cases can be made 
by calculating the probability of finding $a$ in a region between $0$ and $a_{max}$. 
Fig.~\ref{FIG3} shows the quantum probability that the scale factor is smaller than a given value $a$ at 
three different times, for the wave packet with $B_n = 1$, and $k=\pm 1$ (left and right plot respectively)
and also its classical analogue (solid line), using the same phase convention of \cite{MCOFL06}. 
We have used a logarithmic scale in the plots to better appreciate the region around $a=0$.

For $k=1$ we see that at $t=0$ the quantum probability differs strongly from the classical one, since the
scale factor is localized around $a \approx 0.1$, while at later times when the packet spreads, 
the quantum probability gets closer  to its classical counterpart. At $t=0$ only for quite small values of $a$
the classical probability is larger than the quantum one, as a consequence of the Dirichlet boundary conditions
on the wave functions.

More dramatic differences are observed for $k=-1$: in this case the region around $a=0$ is classically 
forbidden and only the quantum probability is nonvanishing around $a = 0$.

Concluding, we feel that the question posed in  \cite{MCOFL06}, whether in the quantum model 
the singularity is avoided, has not been addressed properly in that paper : the observation 
that $\langle a \rangle >0$  is trivial and holds both at classical and quantum level; 
The expectation values calculated for the wave packets considered in  \cite{MCOFL06}
oscillate around the corresponding classical expectation values and can never vanish
neither in classical or in quantum mechanics. The strong oscillations observed for the case 
$k=1$ are understood as a consequence of the Heisenberg principle.

On the other hand, we feel that a better tool to address the question posed by Monerat et al.
is to calculate the probability of finding a scale factor below a given value
(such tool was not considered in \cite{MCOFL06}). Our result show that only for very small $a$
the classical probability is larger than the quantum one, although this effect is a mere consequence
of the Dirichlet boundary conditions, and therefore holds regardless of the binding potential 
used (in this case a quartic potential).

\end{document}